\documentclass[12pt,thmsa]{article}
\usepackage{amsfonts}
\usepackage{graphicx}
\usepackage{epsfig}
\usepackage{graphics}

\makeatletter
\newcommand{\row}[1]%
{\mathord{\buildrel{\lower3pt%
\hbox{$\scriptscriptstyle\rightarrow$}}\over #1}}

\newcommand{\dyadic}[1]{\mathord{\dyadic@rrow{#1}}}
\newcommand{\dyadic@rrow}[1]{
\begin{picture}(12,12)(-1,0)
\put(-1,9){\makebox(0,0)[t]{$\scriptscriptstyle\downarrow$}}
\put(-1,9){\makebox(0,0)[l]{$\scriptscriptstyle\longrightarrow$}}
\put(5,0){\makebox(0,0)[b]{$#1$}}
\end{picture}
}

\newcommand{\bra}[1]{\bigl\langle #1 \bigr|}
\newcommand{\ket}[1]{\bigl| #1 \bigr\rangle}
\newcommand{\expect}[1]{\left\langle #1 \right\rangle}

\begin{document}

\begin{center}
{\Large Unruh acceleration effect on the precision of parameter estimation}\\
 {N. Metwally\\}
$^1$ Department of Mathematics, College of Science, Bahrain
University, Bahrain \\
$^2$Department of Mathematics, Faculty of Science,
Aswan University, Aswan, Egypt \\
\end{center}
\date{\today }

\begin{abstract}
The dynamics of Fisher information for an accelerated system
initially prepared in the  $X$-state  is discussed. An analytical
solution, which consists  of three parts: classical, the average
over all pure states and a mixture of pure states is derived for
the general state and for Werner state. It is shown that, the
Unruh acceleration has a depleting effect on the Fisher
information. This depletion depends on the degree of entanglement
of the initial state settings. For the $X$-state, for some
intervals of Unruh acceleration, the Fisher information remains
constant, irrespective to the Unruh acceleration. In general, the
possibility of estimating the state's parameters decreases as the
acceleration increases. However, the precision of estimation can
be maximized for certain values of the Unruh acceleration. We also
investigate the contribution of the different parts of the Fisher
information on the  dynamics of the total Fisher information.

\end{abstract}
{\bf keywords}: Estimation, Fisher information, Unruh acceleration


\section{\label{sec:Intro}Introduction}

Quantum Fisher information, $\mathcal{F}_I$ represents one of the
most important measures in the context of estimation theory
\cite{Bur}, where it measures the sensitivity of a given state
with respect to changes in one of its parameters. It is considered
as a resource to detect the entanglement between two particles
\cite{Nan}. Quantum Fisher information, is introduced as a
quantitative measure of information flow  between the quantum
system and its surroundings \cite{Lu}. There are several studies
that  have been  devoted to investigate the dynamics of the Fisher
information for different quantum systems. For example, Zhong et.
al \cite{Zhong} have introduced an analytical form of
$\mathcal{F}_I$ for a single qubit in terms of Bloch vectors. The
problem of parameter estimation in a qubit system under different
non-markovian conditions is studied in \cite{Berrada}. Quantum
Fisher information for a system consists of two entangled qubits
interacts with its surrounding is discussed  by Q. Zheng et. al
\cite{Zheng}. Wang et. al \cite{Wang} have investigated the
dynamics of Fisher information for  separable, as well as,
entangled two qubit system, where each qubit interacts
independently with its environment. The dynamics of
$\mathcal{F}_I$ for W-state in the presence  of noisy channel is
examined in \cite{Fatih}. The quantum Fisher information for the
Greenberger-Horne-Zeilinger (GHZ) state in the presence of
decoherence  is investigated by Jian Ma et. al. \cite{Ma}.
Enhancing the quantum teleportation of Fisher information by using
partial measurement is discussed by Xiao et. al.\cite{Xiao}.
Recently an experimental  scheme is proposed by Fr\"{o}wis et.
al\cite{Frowis} to quantify the lower bounds of Fisher
Information.

Investigating the dynamics of Fisher information in non-inertial
frame is  limited. Recently, the effect of the Unruh acceleration
on the  dynamics of $\mathcal{F}_I$ for a single Dirac qubit is
discussed by Banerjee et. al.\cite{Ban}. The dynamics of Fisher
information for a two qubit system in non-inertial frame is
discussed by Yao et. al \cite{Yao}. Here, we are motivated to
study the dynamics of Fisher information for a system of two
qubits initially prepared in the $X$-state or its spacial class
which is known by Werner state. In this current study, we assume
that, only one qubit is accelerated uniformly  while the other
stays in the inertial frame. The Fisher information with respect
to the  output parameters is quantified, where we investigate the
effect of the acceleration and the initial state settings on the
Fisher information.

The paper  is organized as follows: In Sec.$2$, the relations
between Minkowski and Rindler spaces are reviewed and the
mathematical form of the quantum fisher information is introduced.
The suggested model and its dynamics in the Rindler space are
discussed in Sec.$3$, where analytical solutions for the Fisher
information for different parameters are given for the $X$-state
and Werner state.  Finally, a summery and discussion  on the
results are provided in Sec. $4$.

\section{Unruh acceleration and Fisher information}
In the following subsection,(2.1) we review the relation between
inertial and non-inertial frames and effect of Unruh acceleration
on Dirac qubits. In subsection (2.2), we review the main aspects
of quantum Fisher Information and  its mathematical forms.

\subsection{Rindler and Minkowski Spaces}
 Let us assume that, Alice's qubit (subsystem $a$) will be
accelerated, while Bob's qubit (subsystem $b$) will stay in the
inertial frame.  Therefore, if the coordinate of a particle is
defined by $(t,z)$ in Minkowski space, then in Rindler space it is
defined by $(\tau,x)$, where
\begin{equation}
\tau=r \tanh(t/z),\quad x=\sqrt{t^2-z^2},
\end{equation}
and $\quad -\infty<r<\infty, \quad -\infty<x<\infty$,  $\tan
r_b=Exp[-\pi\omega\frac{c}{a_b}]$, ~$0\leq r_b\leq \pi/4$,
$-\infty\leq a\leq\infty$, $\omega$ is the frequency, $c$ is the
speed of  light, and $\phi$ is the phase space which can be
absorbed in the definition of the operators\cite{Jason2013}. These
coordinates define two regions in the Rindler space, $I$ and $II$,
where the accelerated particle will be in the first region $I$ and
the anti-accelerated particle will be in the second region $II$.
In the computational basis $\ket{0_k}$and $\ket{1_k}$  can be
written as\cite{Edu,metwally2016},
\begin{eqnarray}\label{trans}
\ket{0_k}&=&\cos r_a\ket{0_k}_I\ket{0_{-k}}_{II}+ \sin
r_a\ket{1_k}_I\ket{1_{-k}}_{II},\quad
\ket{1_k}=\ket{1_k}_I\ket{0_k}_{II}.
\end{eqnarray}

\subsection{ Fisher Information}
Quantum Fisher information represents the cental role in the
estimation theory, where it can be used  to quantify some
parameters that cannot be quantified directly \cite{Yue}. In this
subsection, we review  the    mathematical form of Fisher
information. Let $\kappa$ be the parameter to be estimated. Let
the spectral decomposition of the density operator is given by
$\rho_{\kappa}=\sum_{j=1}^n\lambda_j\ket{\theta_j}\bra{\theta_j}$,
where $\lambda_j$and $\ket{\theta_j}$ are the eigenvalues and the
corresponding eigenvectors of the state $\rho_\kappa$. The Fisher
information with respect to the parameter $\kappa$ is defined as
\cite{Helstrom}
\begin{equation}
\mathcal{F}_{\kappa}=tr\{\rho_{\kappa}L^2_\kappa\}, ~
L_{\kappa}=\frac{1}{2}(\rho_{\kappa}L_{\kappa}+\L_{\kappa}\rho_{\kappa}),
~L_{\kappa}=\frac{\partial\rho_{\kappa}}{\partial\kappa}.
\end{equation}
Using the spectral decomposition and the identity
$\sum_{j=1}^n\ket{\theta_j}\bra{\theta_j}$ in Eq.(3), one can
obtain the final form of the Fisher information
as\cite{Xiao,Yao,Wei},
\begin{equation}
\mathcal{F}_I=\mathcal{F}_c+\mathcal{F}_p-\mathcal{F}_m,
\end{equation}
where
\begin{eqnarray}\label{Fisher}
\mathcal{F}_{c}&=&\sum_{j=1}^{n}\frac{1}{\lambda_j}\left(\frac{\partial
\lambda_j}{\partial\kappa}\right)^2,
\nonumber\\
\mathcal{F}_p&=& 4\sum_{j=1}^{n}\lambda_j\Bigl(
\expect{\frac{\partial\theta_j}{\partial\kappa}\Big|\frac{\partial\theta_j}{\partial\kappa}}
-\Bigl|\expect{\theta_j\Big|\frac{\partial\theta_j}{\partial\kappa}}\Bigr|^2\Bigr),
\nonumber\\
\mathcal{F}_m&=&8\sum_{j\neq\ell}^{n}\frac{\lambda_j\lambda_\ell}{\lambda_j+\lambda_\ell}
\Big|\expect{\theta_j|\frac{\partial\theta_\ell}{\partial\kappa}}\Big|^2.
\end{eqnarray}
The summations on the first and the third terms over all
$\lambda_j\neq 0$ and $\lambda_j+\lambda_\ell\neq 0$. The first
and the second terms represent the classical Fisher information,
$\mathcal{F}_c$, and the quantum Fisher information,
$\mathcal{F}_q$, respectively. The third term is stemmed from the
mixture of the pure states. Therefore, the quantum  Fisher of a
pure state is larger than that displayed for a mixed state
\cite{Xiao,Yao,Wei}.

\section{ The suggested model}
In this manuscript, the users Alice and Bob share a two-qubit
state of $X$-type \cite{Nasser2013, Nasser2014}. In the
computational basis it can be written as,
\begin{eqnarray}\label{xstate}
 \rho_X&=&\ket{0}_a\bra{0}\Bigl(\mathcal{B}_{11}\ket{0}_b\bra{0}+\mathcal{B}_{22}\ket{1}_b\bra{1}\Bigr)
 +\ket{1}_a\bra{1}\Bigr(\mathcal{B}_{33}\ket{0}_b\bra{0}+
 \mathcal{B}_{44}\ket{1}_b\bra{1}\Bigl)
\nonumber\\
 &&+\ket{0}_a\bra{1}\Bigl(\mathcal{B}_{23}\ket{1}_b\bra{0}+
 \mathcal{B}_{14}\ket{0}_b\bra{1}\Bigr)+
\ket{1}_a\bra{0}\Bigl(\mathcal{B}_{32}\ket{0}_b\bra{1}+
 +\mathcal{B}_{41}\ket{1}_b\bra{0}\Bigr),
 \end{eqnarray}
where
\begin{eqnarray}\label{coef}
\mathcal{B}_{11}&=&\mathcal{B}_{44}=\frac{1}{4}(1+z),\quad\mathcal{B}_{22}=\mathcal{B}_{33}=\frac{1}{4}(1-z),
\nonumber\\
\mathcal{B}_{23}&=&\mathcal{B}_{32}=\frac{1}{4}(x+y),\quad\mathcal{B}_{14}=\mathcal{B}_{41}=\frac{1}{4}(x-y),
\end{eqnarray}
and $x=tr\{\rho_X\sigma_x^{(1)}\sigma_x^{(2)}\}$,
$y=tr\{\rho_X\sigma_y^{(1)}\sigma_y^{(2)}\}$ and
$z=tr\{\rho_X\sigma_z^{(1)}\sigma_z^{(2)}\}$, $\sigma^{j}_k,
j=1,2, k=x,y$ and $z$ are the pauli matrices for the first and the
second qubit, respectively. The subscripts $"a"$ and $"b"$
indicate to Alice and Bob qubits. Since, we are interesting in the
dynamics of the shared state between the users in Rindler space,
it is important to review the relation between Minkowski space and
Rindler space. The next sub-section is devoted for this aim.

\subsection{The $X$-State}

 In this treatment, it is assumed that Alice's qubit
will be accelerated with a uniform acceleration $a$ while Bob's
qubit remains in the inertial frame. By using the initial state
(\ref{xstate}), the transformation (\ref{trans}), and tracing out
the states of  the mode $II$, the final accelerated state between
Alice and Bob can be described as \cite{Nasser2013},
\begin{eqnarray}\label{acc}
\rho_{ab}^{acc}&=&\varrho_{11}\ket{00}\bra{00}+\varrho_{14}\ket{00}\bra{11}+\varrho_{22}\ket{01}\bra{01}
+\varrho_{23}\ket{01}\bra{10} \nonumber\\
&&+\varrho_{32}\ket{10}\bra{01}+\varrho_{33}\ket{10}\bra{10}+\varrho_{41}\ket{11}\bra{00}+\varrho_{44}\ket{11}\bra{11},
\end{eqnarray}
where,
\begin{eqnarray}
\varrho_{11}&=&\mathcal{B}_{11}c^2,\quad
\varrho_{14}=\mathcal{B}_{14}c,\quad
\varrho_{22}=\mathcal{B}_{22}c^2,\quad
\varrho_{23}=\mathcal{B}_{23}c, \nonumber\\
\varrho_{32}&=&\mathcal{B}_{32}c,\quad
\varrho_{33}=\mathcal{B}_{11}s^2+\mathcal{B}_{33},\quad
\varrho_{41}=\mathcal{B}_{41}c,\quad
\varrho_{44}=\mathcal{B}_{22}s^2+\mathcal{B}_{44},
\end{eqnarray}
where $c=\cos r$ and $s=\sin r$. The eigenvalues  of the
accelerated state (\ref{acc}) are given by,
\begin{eqnarray}
\lambda_{1,2}&=&\frac{1}{16}\Bigl\{4+2(1+\cos
2r)z\mp\sqrt{2}\kappa_1(x,y)\Bigr\},
\nonumber\\
\lambda_{3,4}&=& \frac{1}{8}\Bigl\{2-(1+\cos
2r)z\pm2\kappa_2(x,y)\Bigr\},
\end{eqnarray}
where
\begin{eqnarray}
\kappa_1(x,y)&=&\sqrt{4(x-y)^2-4[1-(x-y)^2]\cos 2r+\cos4r+3},
\nonumber\\
\kappa_2(x,y)&=&\sqrt{(x+y)^2\cos^2r+\sin^4r}.
\end{eqnarray}
 and the corresponding eigenvectors are defined as,
\begin{eqnarray}
\ket{\psi_1}=\frac{1}{\sqrt{\mu_1^2+1}}\Bigl(-\mu_1,~0,~0,1\Bigl),
\quad
\ket{\psi_2}=\frac{1}{\sqrt{\mu_2^2+1}}\Bigl(\mu_2,~0,~0,1\Bigl),
\nonumber\\
\ket{\psi_3}=\frac{1}{\sqrt{\mu_3^2+1}}\Bigl(0,~-\mu_3,~1,0\Bigl),
\quad
\ket{\psi_4}=\frac{1}{\sqrt{\mu_4^2+1}}\Bigl(0,~\mu_4,~1,0\Bigl),
\end{eqnarray}
where
\begin{eqnarray}
\mu_{1,2}&=&\pm\frac{\sec r}{4(x-y)}\Bigl\{2(1-\cos2r)\pm\sqrt{2}\kappa_1(x,y))\Bigr\}, \nonumber\\
\mu_{3,4}&=&\pm\frac{\sec
r}{4(x+y)}\Bigl\{2(1-\cos2r)\pm\sqrt{2}\kappa_3(x,y))\Bigr\},
\end{eqnarray}
with $\kappa_3(x,y)=\sqrt{4(x+y)-4[1-(x+y)^2]\cos 2r+\cos 4r+3}$.

It is evident that, the final accelerated state (8) is a function
of the parameters $x,y,z$ and $r$. Therefore, we can estimate
these parameters  by means of Fisher information. In what follows,
we calculate $\mathcal{F}^z_{I}$, $\mathcal{F}^x_{I}$ and
$\mathcal{F}^r_{I}$.

\begin{itemize}
\item {\it Fisher information with respect to $z$, $\mathcal{F}^z_{I}$}:\\
It is noted that, the eigenvectors don't depend on the parameter
$z$, therefore, there is no any contribution from $\mathcal{F}_q$
and $\mathcal{F}_m$. However,  the total Fisher information with
respect to the parameter $z$ depends on the classical part, which
can be written explicitly as
\begin{eqnarray}
\mathcal{F}^z_I&=&\frac{(1+\cos2r)^2}{4}\Bigr[\frac{8+4(1+\cos2r)z}{(4+2(1+\cos2r)z)^2-2\kappa^2_1(x,y)}
\nonumber\\
&&\hspace{5cm}+
\frac{2-(1+\cos2r)z}{(2-(1+cos2r)z)^2-4\kappa^2_2(x,y)}\Bigr].
\end{eqnarray}

\item{\it  Fisher information with respect to $x$,
$\mathcal{F}_{I}^x$:}\\
Using Eqs.(10-13) together with (5), we can write  the components
of the Fisher information as,
\begin{eqnarray}
\mathcal{F}^x_{c}&=&\frac{(x-y)(1+\cos
2r)^2}{\kappa^2_1(x,y)}\Bigl\{\frac{2+(1+\cos 2r)z}{[4+2(1+\cos
2r)z]^2-2\kappa^2_1(x,y)}\Bigr\}
\nonumber\\
&&+ \frac{(x+y)^2\cos^4 r}{\kappa^2_2(x,y)}\Bigl\{\frac{2-(1+\cos
2r)z}{[2-2(1+\cos 2r)z]^2-4\kappa^2_2(x,y)}\Bigr\},
\nonumber\\
\mathcal{F}^x_p&=&4\sum_{i=1}^4\lambda_i\frac{\mu'^2_i}{(1+\mu_i^2)^2},
\quad i=1...4, ~\mu'_i=\frac{d\mu_i}{dx},
 \nonumber\\
 \mathcal{F}_m&=&\frac{8\lambda_1\lambda_2}{(\lambda_1+\lambda_2)}\frac{(\mu_1+\mu_2)^2}{(1+\mu_1^2)(1+\mu_2^2)}
 \Bigl\{\frac{\mu_1'^2}{(1+\mu_1^2)^2}+\frac{\mu_2'^2}{(1+\mu_2^2)^2}\Bigr\}
 \nonumber\\
 &&+\frac{8\lambda_3\lambda_4}{(\lambda_3+\lambda_4)}\frac{(\mu_3+\mu_4)^2}{(1+\mu_3^2)(1+\mu_4^2)}
 \Bigl\{\frac{\mu_3'^2}{(1+\mu_3^2)^2}+\frac{\mu_4'^2}{(1+\mu_4^2)^2}\Bigr\}.
\end{eqnarray}

\begin{figure}
\centering
           \includegraphics[width=21pc,height=16pc]{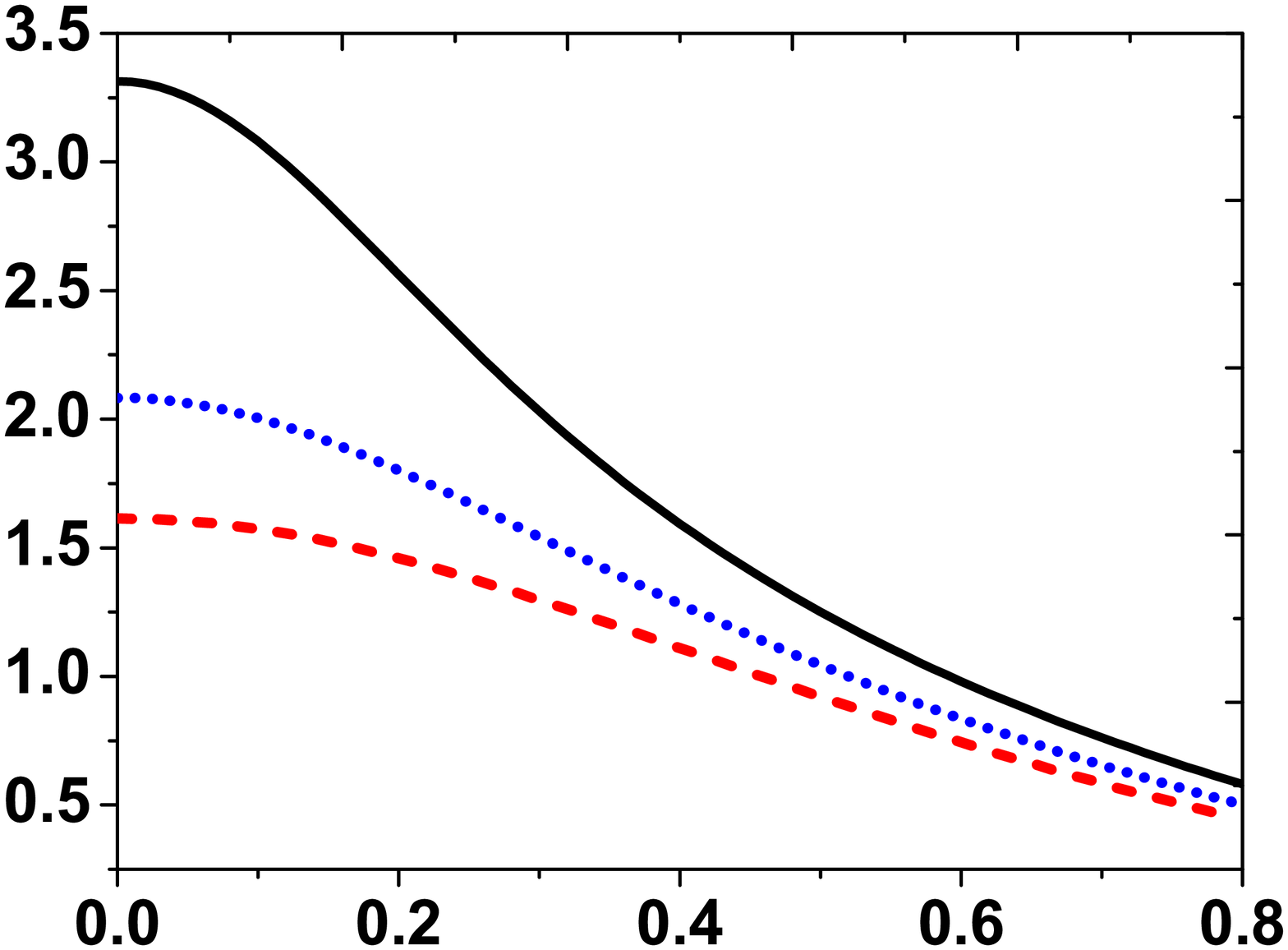}
             \put(-130,7){\Large$r$}
     \put(-250,90){\Large$\mathcal{F}^z_I$}
     \put(-70,155){$(a)$}
       \includegraphics[width=21pc,height=16pc]{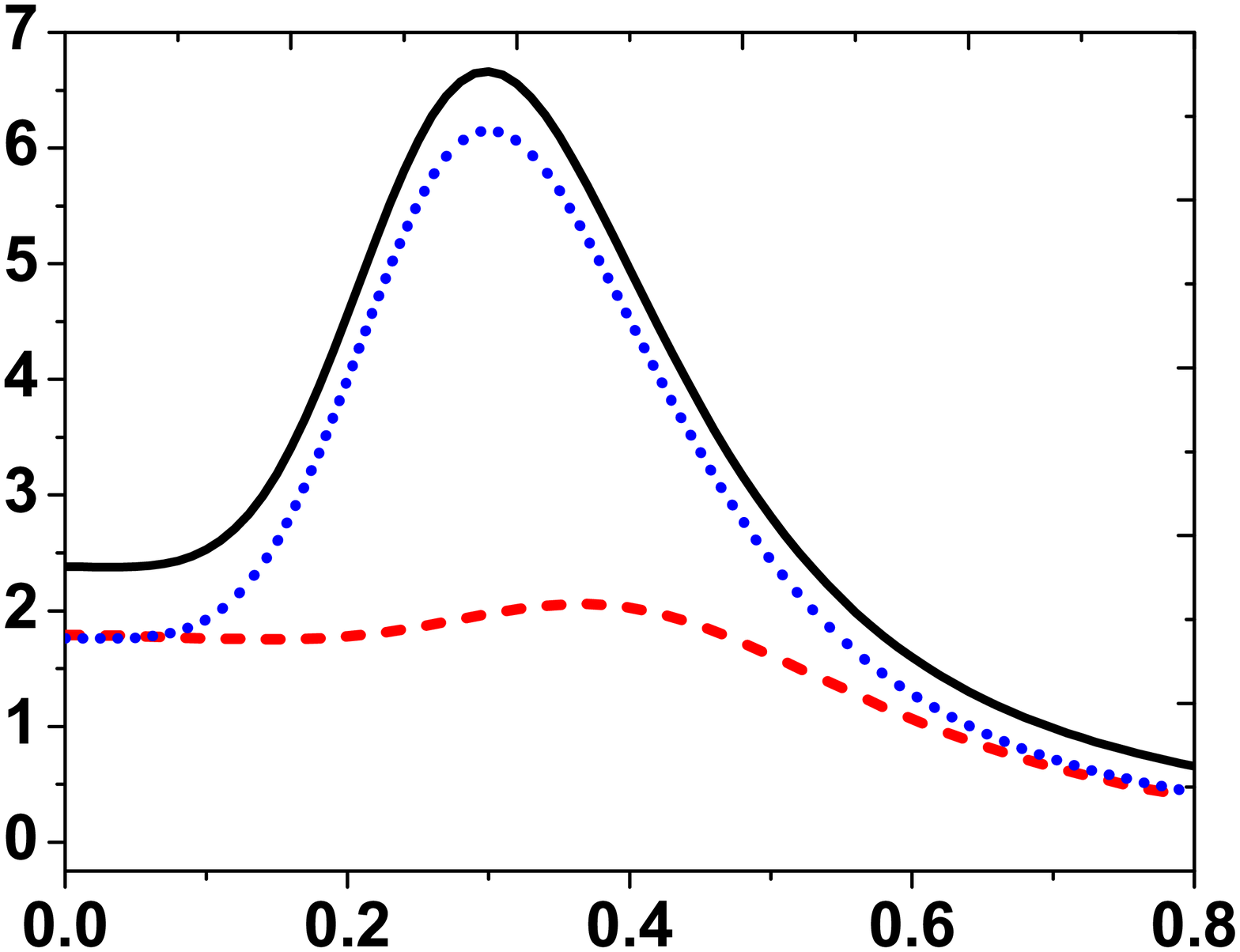}
              \put(-130,7){\Large$r$}
     \put(-250,90){\Large$\mathcal{F}^x_I$}
      \put(-70,155){$(b)$}
            \caption{(a)~Fisher information, $\mathcal{F}^z_{I}$ for an accelerated system initially prepared
            the X-state with $x=-0.3,y=-0.6$ and
            $z=-0.3,-0.5,-0.6$ dash-dot and solid curves, respectively.
       (b) Fisher information with respect to $x$, $\mathcal{F}^x_{I}$,
       where we set $y=-0.6, z=-0.5$ and $x=-0.4,-0.5$ and
       $-0.7$ for the dash, dot and solid curves, respectively.  }
\end{figure}

The dynamics of Fisher information with respect to the parameters
$z$ and $x$  is shown in Fig.(1).  Different initial state
settings of $X-$ state are considered. Fig.(1a) shows the behavior
of $\mathcal{F}^z_{I}$ for different values of $z$, where we fix
the values of $x(=-0.3)$ and $y(=-0.6)$. The general behavior
shows that the Fisher information decreases as the acceleration
increases. The decreasing rate depends on the initial state
settings, where as one increases $|z|$, the upper bounds of Fisher
information increase. This means that as one increases the
entanglement of the initial state, the Fisher information with
respect to $z$ increases.

The behavior of $\mathcal{F}^x_{I}$ for different initial state
settings is depicted in Fig.(1b). There are a  three behaviors
manifested   of Fisher information: for small values of the
acceleration, Fisher Information $\mathcal{F}^x_{I}$ remains
constant. For further values of $r$, the  Fisher information with
respect to $x$, increases suddenly to reach its maximum value at a
certain finite value of $r$. However for larger values of the
acceleration, $\mathcal{F}^x_{I}$ decreases gradually to reach its
minimum values at $r\to\infty$. The upper and lower values of
$\mathcal{F}^x_{I}$ depend on the initial value of the parameter
$x$, where as one increases $|x|$, the upper bounds increase.

Generally speaking,  larger acceleration leads to smaller values
of the Fisher information, namely, less precision of estimation on
the $z$ and $x$ parameters. The possibility of estimating these
parameters increases for systems initially that has a large degree
of entanglement. Another feature is depicted, Fisher information
with respect to $x$ has a maximum value at a certain value of the
acceleration, while that with respect to $z$ decreases
monotonously with respect to $r$.

\begin{figure}
\centering
           \includegraphics[width=25pc,height=16pc]{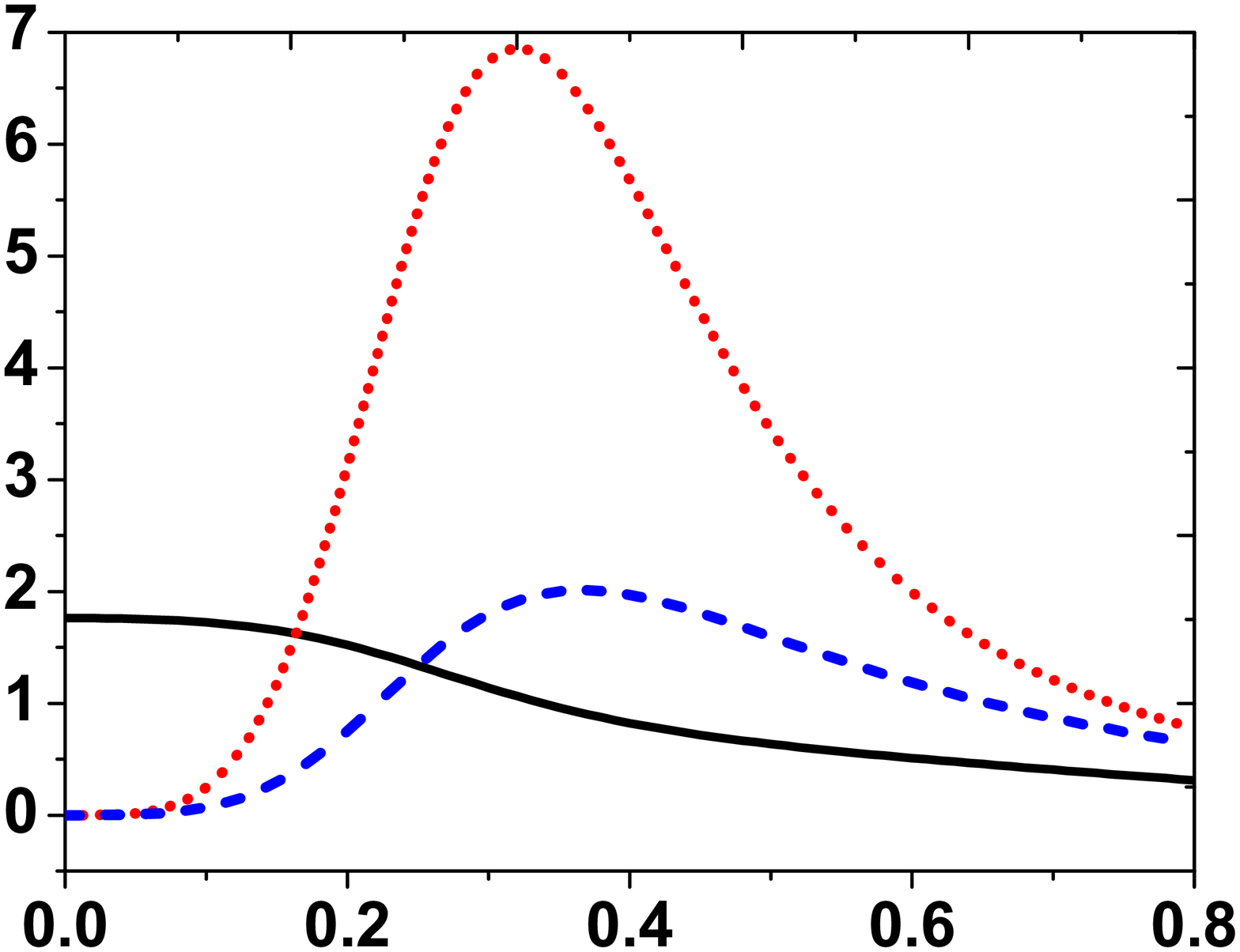}
             \put(-150,7){\Large$r$}
     \put(-285,95){\Large$\mathcal{F}^x_I$}
                \caption{The components of Fisher information $\mathcal{F}_{I}^x$   for a system is initially
            prepared in $X$-state with $x=-0.5,y=-0.6$ and $z=-0.5$. The solid, dot and dash curves for $\mathcal{F}^x_c,
            \mathcal{F}^x_p$ and $\mathcal{F}^x_m$, respectively.}
\end{figure}

In Fig.(2), we plot the three different type of Fisher
information, $\mathcal{F}^x_c,~ \mathcal{F}^x_p$ and
$\mathcal{F}^x_m$, where it is assumed that the system is
initially prepared in the $x-$ state with $x=-0.5,y=-0.6$ and
$z=-0.5$. It is evident that, at $r=0$, the total Fisher
information, $\mathcal{F}_{I}^x=\mathcal{F}^x_c$, while there is
no any contribution from $\mathcal{F}^x_p$ and $\mathcal{F}^x_m$.
As the Unruh acceleration increases, the classical component of
Fisher information decreases, while the pure and mixed parts
increase. However each of $\mathcal{F}^x_p$ and $\mathcal{F}^x_m$
reaches their maximum values at a certain value of the Unruh
acceleration. This displays that, there are some  pure states that
are generated at $r\in[0.1,0.4]$. For further values of
$r\in[0.4,0.6]$, the number of the generated pure states
decreases.

\begin{figure}
\centering
            \includegraphics[width=30pc,height=20pc]{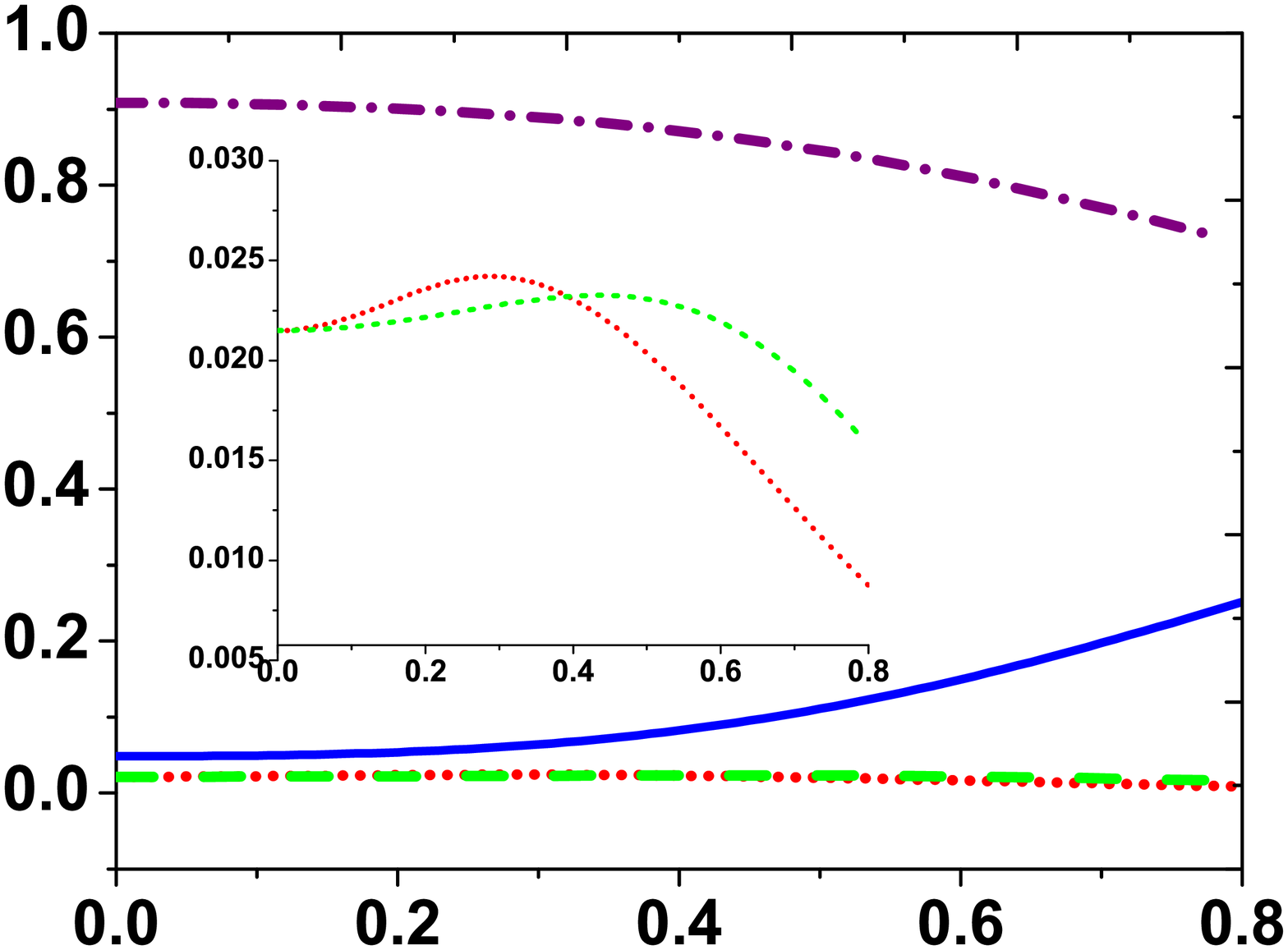}
              \put(-200,20){\Large$r$}
    \put(-360,130){\Large$\mathcal{P}_{ix}$}
            \caption{The dynamics of the populations, $\mathcal{P}_{\psi_{1x}}$(dot-curve),~
            $\mathcal{P}_{\psi_{2x}}$(solid curve) ,
            $\mathcal{P}_{\psi_{3x}}$(dash curve), $\mathcal{P}_{\psi_{4x}}$(dash-dot curve). The small scale Figure
            display the behavior of $\mathcal{P}_{\psi_{1x}}$ and $\mathcal{P}_{\psi_{3x}}$.}
\end{figure}

To explain the behavior of the Fisher information, it is important
to plot the behavior of the probabilities distribution of the
different eigenvectors, where
$\mathcal{P}_{ix}=\lambda_i^2/\lambda$,
$\lambda=\sum_{i=1}^{4}\lambda^2_i$. Therefore, we consider the
case for $\mathcal{F}_I^x$ with $x=-0.5,~y=-0.6$ and $z=-0.5$.
Fig.(3) describes the behaviors of $\mathcal{P}_i, i=1..4$. It is
evident that, the travelling state can be written in a different
structure in the interval $r\in[0,~0.8]$. For example for
$r\in[0,0.05]$, the final density operator can be written as,
$\rho_1=\kappa_1(\ket{\psi_{1x}}\bra{\psi_{1x}}+\ket{\psi_{1x}}\bra{\psi_{1x}})+
\kappa_2\ket{\psi_{2x}}\bra{\psi_{2x}}+\kappa_3\ket{\psi_{4x}}\bra{\psi_{4x}}$.
However, for $r\in[0.05,0.4)$ another form can be written as
$\rho_2=\sum_{j}\kappa_j\ket{\psi_{ix}}\bra{\psi_{jx}},j=1..4$.
For $r=0.4$ a third structure can be depicted of the travelling
state, $\rho_3$ which is similar to that shown for $\rho_1$ but
with different values of $\kappa_j$. For further values of
$r\in(0.4,0.8]$ the travelling state can be written as $\rho_2$
but with a different $\kappa_j$'s. From these forms, it is seen
that the possibility of generating pure states is large. This
explain why the contribution from $\mathcal{F}_{p}^x$ is the
largest one.

\item{\it Fisher information with respect to r,
$\mathcal{F}_I^r$:}\\
In this case, the classical part of Fisher information
$\mathcal{F}^r_c$ is given by
\begin{eqnarray}
\mathcal{F}_c^r&=&\frac{1}{2\lambda_1\kappa_1^2(x,y)}\Bigl[\sqrt{2}z\kappa_1(x,y)\sin2r+2[1-(x-y)^2]\sin2r-\sin4r\Bigr]^2
\nonumber\\
&&+\frac{1}{2\lambda_2\kappa_1^2(x,y)}\Bigl[\sqrt{2}z\kappa_1(x,y)\sin2r-2[1-(x-y)^2]\sin2r+\sin4r\Bigr]^2
\nonumber\\
&&+\frac{\sin2r^2}{\lambda_3\kappa^2_2(x,y)}\Bigl[(x+y)^2-2\sin^2r+2z\kappa_2(x,y)\Bigr]^2
\nonumber\\
&&+\frac{\sin2r^2}{\lambda_4\kappa^2_2(x,y)}\Bigl[(x+y)^2-2\sin^2r-2z\kappa_2(x,y)\Bigr]^2.
\end{eqnarray}
The other components $\mathcal{F}_p^r$ and $\mathcal{F}_m^r$ are
similar to those shown in Eqs.(15) but the derivative will be with
respect to the parameter $r$, namely, we replace
$\frac{\partial\mu_i}{\partial x}$ by
$\frac{\partial\mu_i}{\partial r}$.
\begin{figure}
\centering
           \includegraphics[width=25pc,height=16pc]{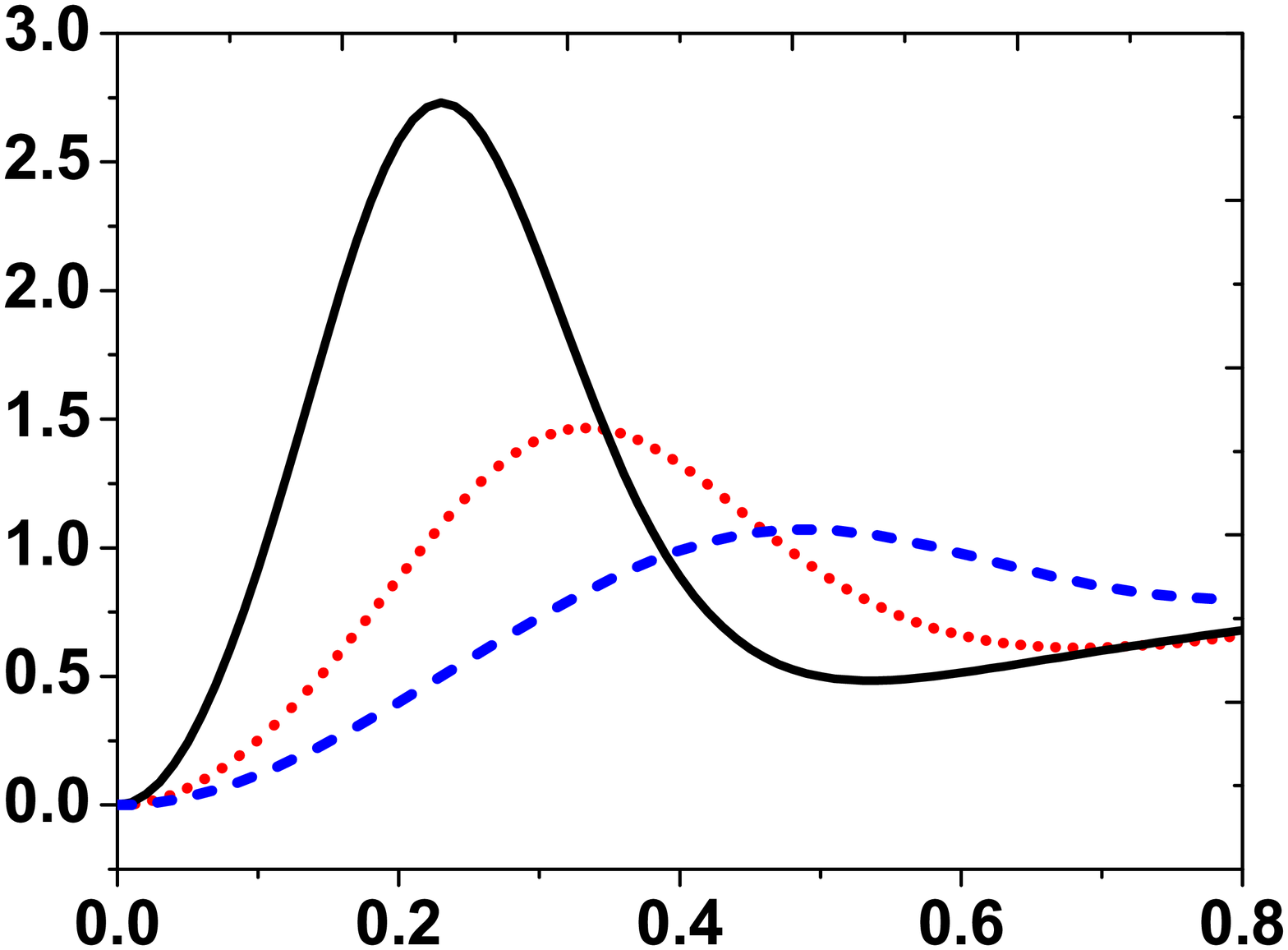}
           \put(-150,7){\Large$r$}
     \put(-290,95){\Large$\mathcal{F}^r_I$}
                 \caption{Fisher Information, $\mathcal{F}^r_{I}$ for an accelerated system initially prepared
            the X-state with $y=-0.6,z=-0.5$ and
            $x=-0.2,-0.4,-0.5$ dash, dot and solid  curves, respectively. }
\end{figure}

The dynamics of Fisher information with respect to the Unruh
acceleration  parameter $\mathcal{F}_I^r$ is displayed in Fig.(4).
Different initial state settings are considered. The behavior
shows that at $r=0$, $\mathcal{F}^r_I=0$. As soon as the particle
is accelerated, the Fisher information increases as Unruh
acceleration increases sharply for systems that possess large
degree of entanglement. The Fisher information reaches their
maximum values at  a certain value of acceleration. An important
remark that,  systems that possess a small degree of entanglement
reaches their upper bounds at larger values of Unruh acceleration.
For further values of $r$, the  Fisher information,
$\mathcal{F}_I^r$  decays faster for systems that have a large
degree of entanglement. As $r\to\infty$, the upper bounds of
Fisher information  depends on the degree of entanglement of the
initial accelerated systems.

From this figure, it is seen  that, it is possibly to estimate the
Unruh acceleration parameter with high precision if the
accelerated systems have a large degree of entanglement at smaller
accelerations.
\end{itemize}

\subsection{Werner state}
Here, we consider  the users share initially a two qubit state of
Werner type $(\rho_w)$. This means that, we set $x=y=z$ in
(\ref{coef}). According to the previous suggestion i.e., only
Alice's qubit will be accelerated, then at the end of the process
the users will obtain an accelerated Werner state $\rho_w^{acc}$
with the following eigenvalues,
\begin{eqnarray}\label{Ew}
\lambda_{1w}&=&\frac{c^2}{4}(1+x),\quad
\lambda_{2w}=\frac{1}{4}(1+s^2+c^2x), \nonumber\\
\lambda_{3w,4w}&=&\frac{1}{4}\Bigl\{1-x\cos^2
r\mp\frac{1}{2\sqrt{2}}\sqrt{16x^2-4(1-4x^2)\cos 2r+\cos 4r
+3}\Bigr\},
\end{eqnarray}
and their corresponds eigenvectors,
\begin{eqnarray}\label{Vw}
\ket{\psi_{1w}}&=&\Bigl(1,~0,~0,~0\Bigr),\quad
\ket{\psi_{2w}}=\Bigl(0,~0,~0,~1\Bigr), \nonumber\\
\ket{\psi_{3w}}&=&\frac{1}{\sqrt{1+\mu_{3w}^2}}\Bigl(0,~-\mu_{3w},~1,~0\Bigr),
\ket{\psi_{4w}}=\frac{1}{\sqrt{1+\mu_{4w}^2}}\Bigl(0,~-\mu_{4w},~1,~0\Bigr),
\end{eqnarray}
where $\mu_{3w,4w}=\frac{1}{8x\cos r}\Bigl[(2-2\cos
2r+\gamma\Bigr]$, $\gamma=\sqrt{(16x^2-4(1-4x^2)\cos2r+\cos4r+3}$.

\begin{itemize}
\item{Estimating $x$-parameter:\\}
 Now, we have all details to evaluate Fisher information for
the parameter $x$ and the parameter $R$. Inserting  (\ref{Ew}),
(\ref{Vw}) in the definition of Fisher information (\ref{Fisher}),
one gets
$\mathcal{F}^w_{xI}=\mathcal{F}^w_{xc}+\mathcal{F}^w_{xp}-\mathcal{F}^w_{xm}$
explicitly where,

\begin{figure}
\centering
           \includegraphics[width=21pc,height=16pc]{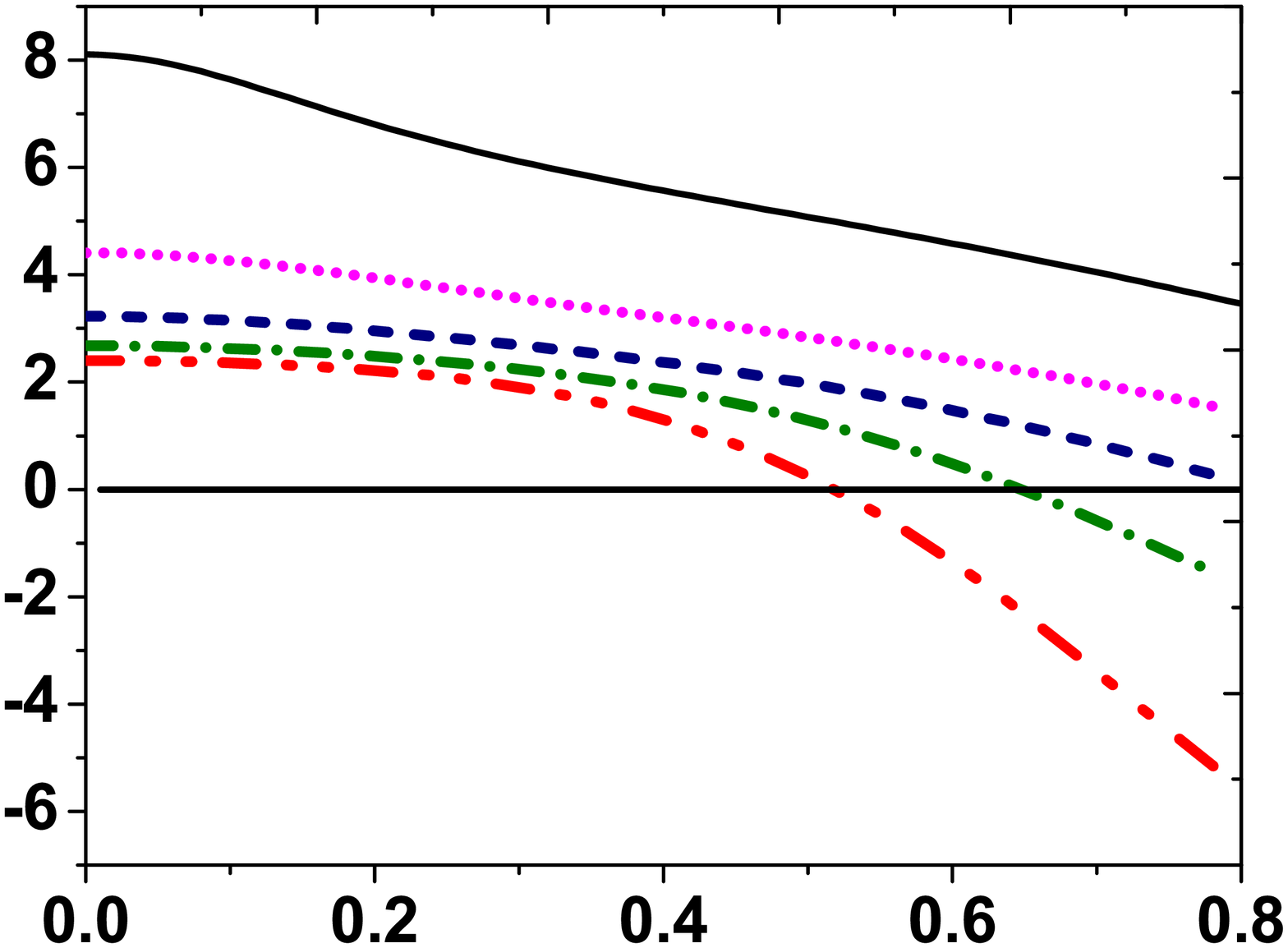}
             \put(-130,7){\Large$r$}
     \put(-250,90){\Large$\mathcal{F}^w_{xI}$}
     \put(-70,155){$(a)$}
       \includegraphics[width=21pc,height=16pc]{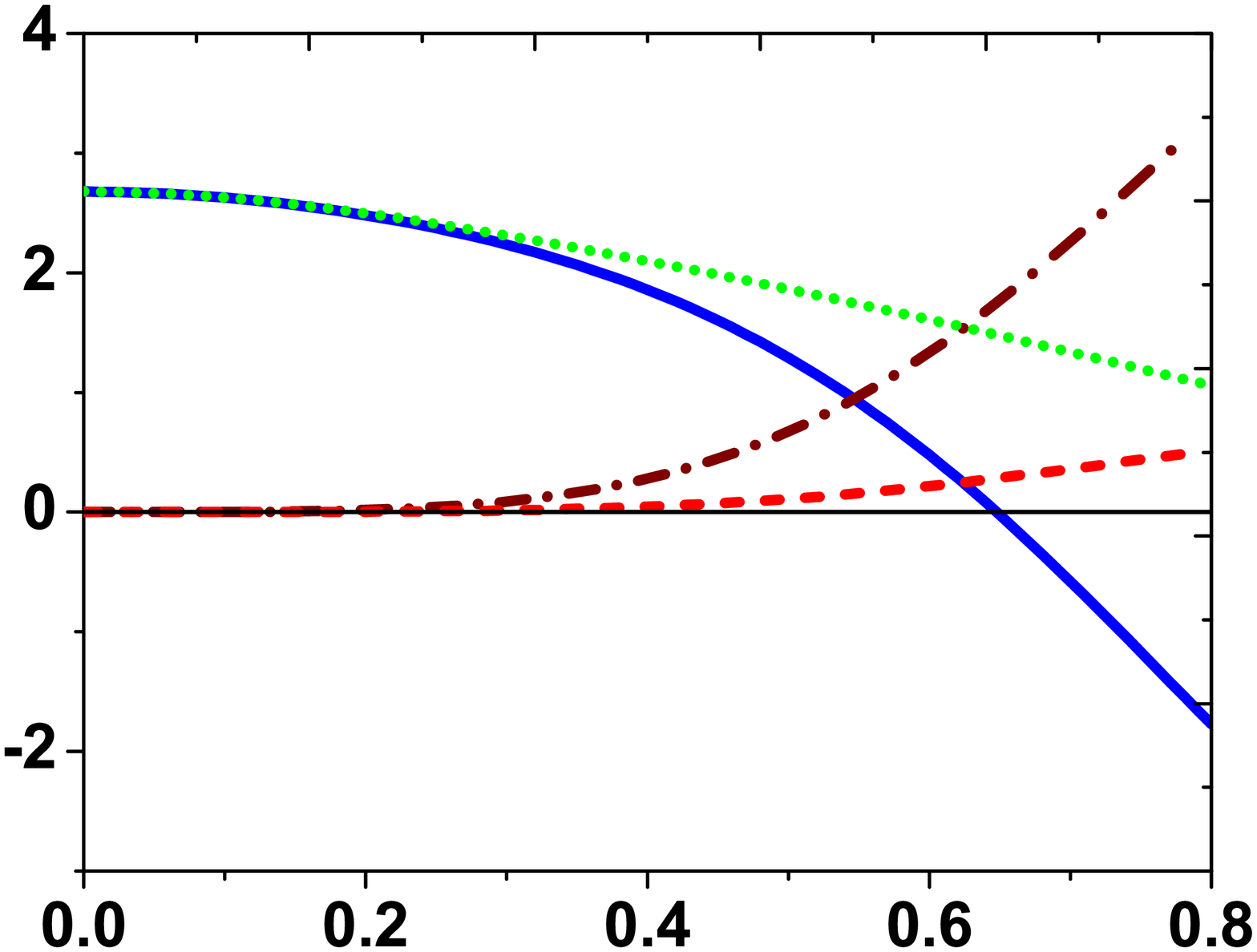}
              \put(-130,7){\Large$r$}
     \put(-250,90){\Large$\mathcal{F}^w_{xJ}$}
     \put(-70,155){$(b)$}
            \caption{(a)Fisher information, $\mathcal{F}^w_{xI}$ for an accelerated system initially prepared in Werner states.
       The solid, dot, dash, dash-dot, and dash-dot-dot curves for $x=-0.9,-0.8,-0.7, -0.6,-0.5$, respectively.
       (b) $\mathcal{F}^w_{xJ}$ represents the contributions from the classical $\mathcal{F}^w_{xc}$(dot curve),
              the pure part, $\mathcal{F}^w_{xp}$ (dash-curve)and  from the mixed part $\mathcal{F}_{xm}$(dash-dot curve),
               while the solid curve represents the total Fisher information, $\mathcal{F}^w_{xI}$.  }
\end{figure}

\begin{eqnarray}\label{FisherW}
\nonumber\\
\mathcal{F}^w_{xc} &=&\frac{\cos^2 r}{4(1+x)}+\frac{\cos^4
r}{4(1+\sin^2r+x\cos^2r)} + \frac{\Bigl(\gamma\cos^2r+8x(1+\cos
2r)\Bigr)^2}{\gamma^2\left(1-x\cos^2r-\frac{1}{2\sqrt{2}}\gamma\right)}
\nonumber\\
&&+ \frac{\Bigl(\gamma\cos^2r-8x(1+\cos
2r)\Bigr)^2}{\gamma^2\left(1-x\cos^2r+\frac{1}{2\sqrt{2}}\gamma\right)},
\nonumber\\
\mathcal{F}^w_{xp}&=&
4\Bigl[\lambda_{3w}\frac{\mu'^2_{3w}}{(1+\mu^2_{3w})^2}+\lambda_{4w}
\frac{\mu'^2_{4w}}{(1+\mu^2_{4w})^2}\Bigr],
\nonumber\\
\mathcal{F}^w_{xm}&=&\frac{8\lambda_{3w}\lambda_{4w}}{\lambda_{3w}+\lambda_{4w}}\Bigl[\frac{(\mu_{3w}+\mu_{4w})^2}{(1+\mu^2_{3w})(1+\mu^2_{4w})}\Bigr]
\Bigl\{\frac{\mu'^2_{4w}}{(1+\mu^2_{4w})^2}
+\frac{\mu'^2_{3w}}{(1+\mu^2_{3w})^2},
 \Bigr\},
\end{eqnarray}
where ,
\begin{eqnarray}
\mu'_{3w}&=&\frac{\partial{\mu_{3w}}}{\partial x}=-\frac{1}{\cos
r}\Bigl[\frac{4}{\sqrt{2}\gamma}(1+\cos
2r)-\frac{2-2\cos2r+\sqrt{2}\gamma}{8x^2}\Bigr],
\nonumber\\
\mu'_{4w}&=&\frac{\partial{\mu_{4w}}}{\partial x}=\frac{1}{\cos
r}\Bigl[\frac{4}{\sqrt{2}\gamma}(1+\cos
2r)-\frac{2-2\cos2r-\sqrt{2}\gamma}{8x^2}\Bigr].
\end{eqnarray}

 In Fig.(5a), we showed the behavior of the Fisher information ($\mathcal{F}^w_{xI}$ for a
 system  initially prepared in a two qubit state of Werner
 type). This state affect by Unruh effect, where only the first qubit (Alice'squit) undergos a uniform acceleration, while
 the second qubit (Bob's qubit) remains  in an inertial frame. Due to
 this effect, the Fisher information decreases as the acceleration
 increases. The decay rate of $\mathcal{F}^w_{xI}$ depends on the
 initial state settings and the acceleration value. It is clear
 that, for higher entangled state, the Fisher information is
 larger than that depicted for less entangled states. For small
 value of acceleration, $\mathcal{F}^w_{xI}$ is almost a stable.

From the definition of the Fisher information, it consists of
three parts. One part represents the classical contribution, the
second one represents the contributions of all pure states and the
third  part represents  the mixture of all pure states. In
Fig.(5b), we investigate the behavior of the three types of these
information where we assume that the system is initially prepared
in a Werner state with $x=-0.6$.

\begin{figure}[b!]
\centering
           \includegraphics[width=21pc,height=16pc]{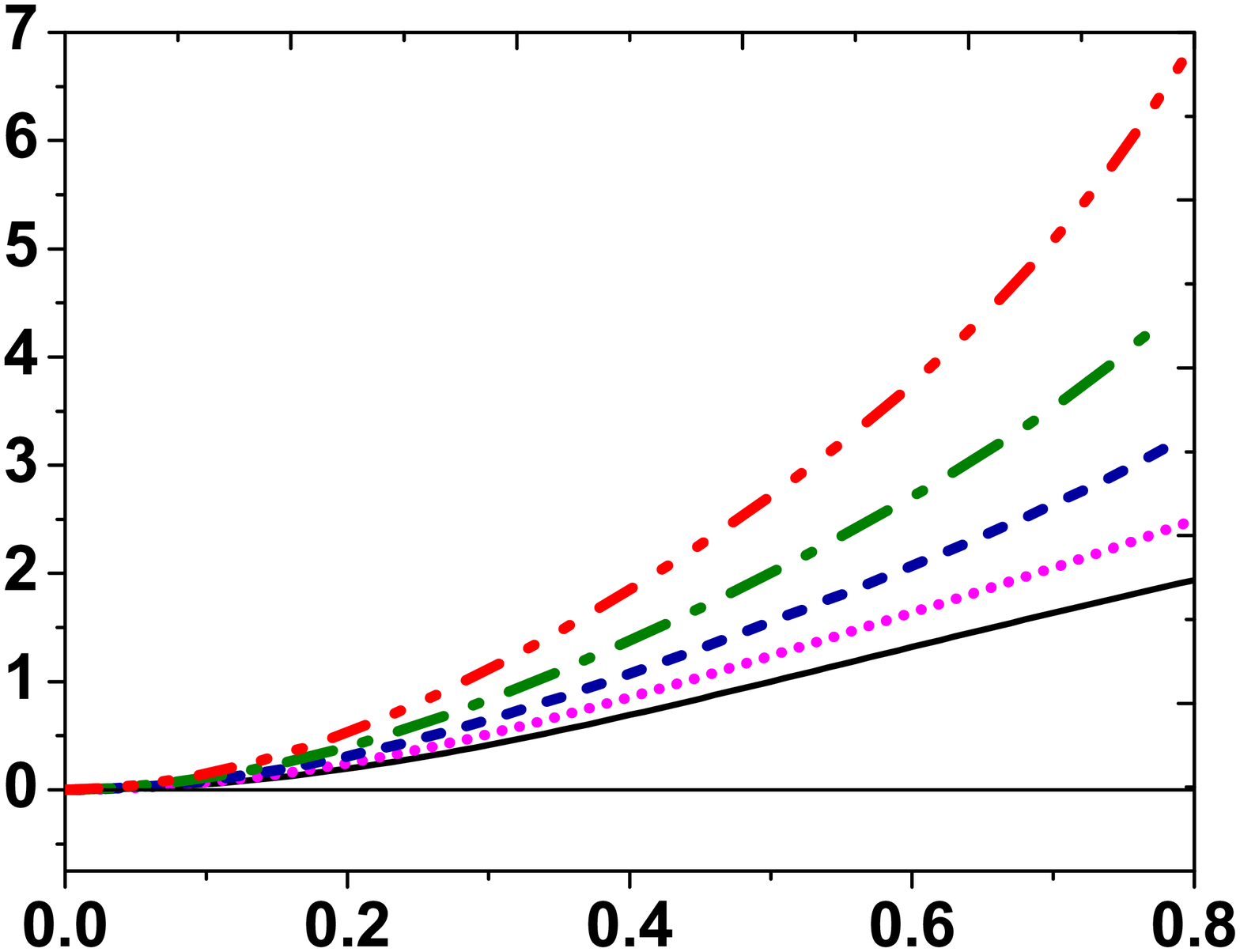}
           \put(-130,7){\Large$r$}
     \put(-250,90){\Large$\mathcal{F}^w_{rI}$}
     \put(-70,155){$(a)$}
       \includegraphics[width=21pc,height=16pc]{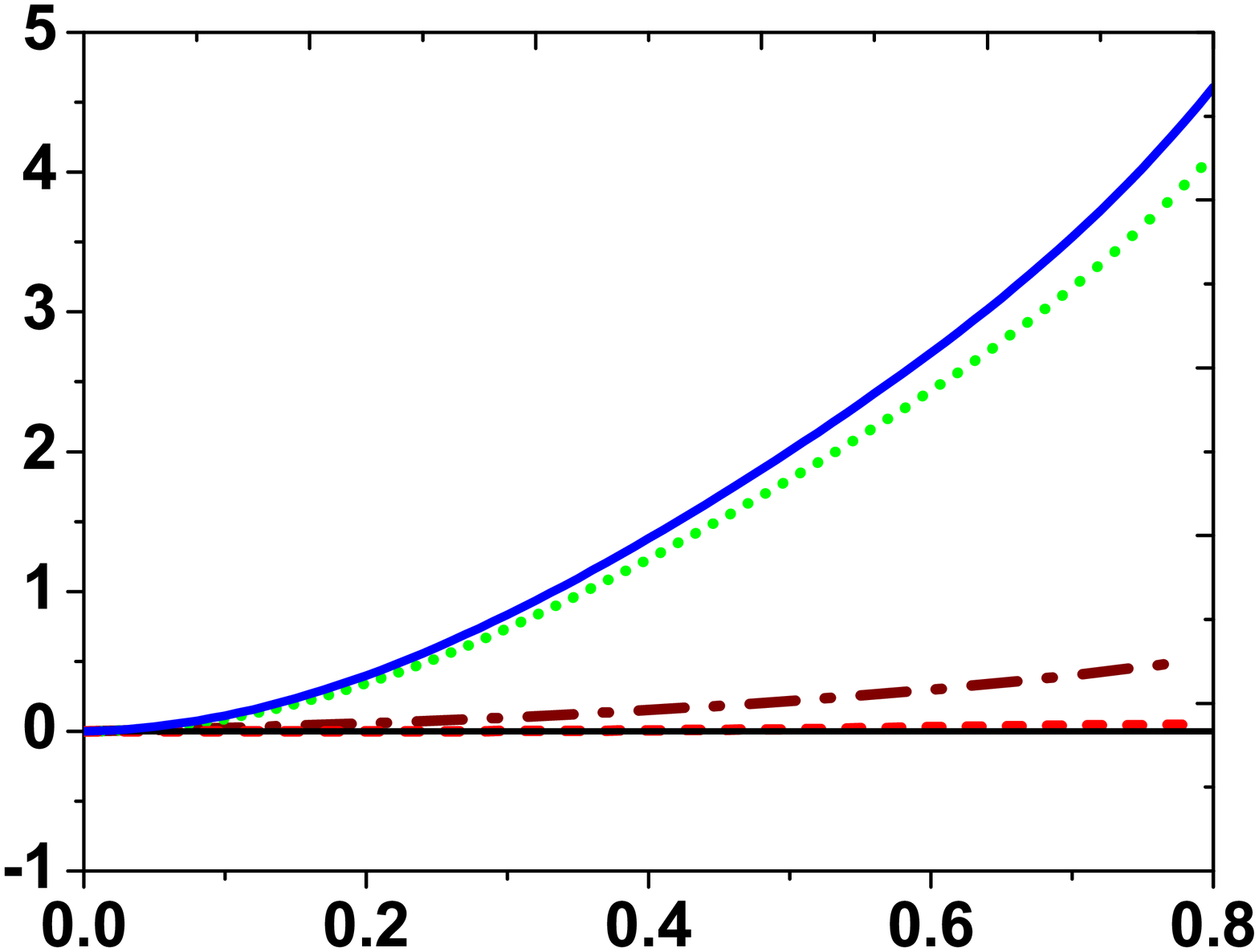}
            \put(-130,7){\Large$r$}
     \put(-250,90){\Large$\mathcal{F}^w_{rJ}$}
     \put(-70,155){$(b)$}
            \caption{Fisher information, $\mathcal{F}^w_{rI}$ for an accelerated system initially prepared in Werner states.
      (a) The solid, dot, dash, dash-dot, and dash-dot-dot curves for $x=-0.9, -0.8, -0.7, -0.6$, respectively.
      (b) The total Fisher information $\mathcal{F}^w_{rI}$ (solid curve)
       $\tilde\mathcal{F}_c$ (dot curve), $\mathcal{F}^w_{rp}$
      (dash-curve) and $\mathcal{F}^w_{rm}$ (dash-dot curve),
      where $x=-0.6$.}
\end{figure}

\item{Estimating the acceleration parameter, $r$:\\} Using the
eigenvalues (\ref{Ew}) and the normalized eigenvectors (\ref{Vw})
in (\ref{Fisher}), one obtains a similar expressions to that given
by (\ref{FisherW}), where
$\mathcal{F}^W_{rI}=\mathcal{F}^w_{rc}+\mathcal{F}^w_{rq}-\mathcal{F}^w_{rm}$.
The expressions of $\mathcal{F}^w_{rq} $ and  $\mathcal{F}^w_{rm}$
are similar to that obtained in (\ref{FisherW}), but the
derivatives wit respect to the parameter $r$. Namely we replace
$\frac{\partial{\mu_{3w}}}{\partial x}$  and
$\frac{\partial{\mu_{4}w}}{\partial x}$ by
$\frac{\partial{\mu_{3}w}}{\partial r}$ and
$\frac{\partial{\mu_{4w}}}{\partial r}$, respectively, where

\begin{eqnarray}
\frac{\partial{\mu_{iw}}}{\partial r}&=&\pm\frac{\sec
r}{2x}\Bigl[\sin2r\pm\frac{1}{\sqrt{2}\gamma}\left(2(1-4x^2)\sin2r-\sin4r\right)
\nonumber\\
&&\hspace{5cm} \mp \frac{1}{2}(1-\cos
2r)-\frac{1}{2\sqrt{2}}\gamma\tan r\Bigr],
\end{eqnarray}
where $i=3,4$, respectively.

 While the
expression of $\mathcal{F}^r_m$ takes the following form,
\begin{eqnarray}
\mathcal{F}^w_{rc}&=&(1+x)\sin^2r+\frac{1-x}{4}\sin2r+\frac{1}{2\gamma^2}\Bigl[\frac{\left(\sqrt{2}x\gamma\sin
2r+2(1-4x^2)\sin2r+\sin 4
r\right)^2}{4-4x\cos^2r-\sqrt{2}\gamma}\Bigr] \nonumber\\
&&+\frac{1}{2\gamma^2}\Bigl[\frac{\left(\sqrt{2}x\gamma\sin
2r+2(1-4x^2)\sin2r-\sin 4
r\right)^2}{4-4x\cos^2r+\sqrt{2}\gamma}\Bigr].
\end{eqnarray}

In Fig.(6a), the behavior of $\mathcal{F}^w_{rI}$ is investigated
for different initial state settings. It is clear that, the Fisher
information increases as the acceleration increases. The
increasing rate depends on the initial entanglement of the
accelerated state. However, for less initial entanglement state,
the increasing rate of $\mathcal{F}^w_{rI}$ is larger than that
displayed for higher entangled state. This shows that we can
measure the parameter $r$ with high precision

Fig.(6b) shows the contributions from the different types of the
Fisher information, where we assume that the system is initially
prepared in Werner state described by $x=-0.6$. It is clear that
all $\mathcal{F}^w_{rj}, ~J=c,p,m$
 increase as $r$ increases, but with a different increasing rate.
 However, the contribution from the classical part is the largest
 one, while the pure part, $\mathcal{F}^w_{rp}$ represent the smallest
 contribution. This shows that the possibility of obtaining
 classical correlated system is very large. On the other hand, this possibility decreases for obtaining pure
 states.

\end{itemize}

\section{Conclusion}
The behavior of Fisher information with respect to the parameters
of an accelerated $X$-state is discussed, where we assume only one
particle is  uniformly accelerated  and the other partial remains
in the initial frame. A detailed exposition of the analytical form
of the Fisher  information is provided for the general state and
its special version (Werner state). The effect of the Unruh
acceleration on the dynamics of the Fisher information is
investigated for different initial state settings. This study
helps in revealing the  sensitivity of the suggested system  to
the change of the state parameters. The contribution of the
components of the Fisher information, classical, average pure and
mixture on the total Fisher information  is inspected.

The results show that, Fisher information  has different behaviors
depending on the initial state settings and the estimated
parameter. In general, Fisher information  decays as the Unruh
acceleration increases. This decay may be steeper or gradually
depending on the value of the acceleration and the values of the
initial  parameters. Three different behaviors are depicted for
the Fisher information  namely, constant, sudden increasing and
sudden decay. The upper bounds of Fisher information are large for
systems  that have  large degree of entanglement and consequently
the precision of estimating the parameters increases. On the other
hand, as the acceleration goes to infinity the lower bounds of the
Fisher information are non zero.

Fisher information is quantified  with respect to the $z, x$ and
$r$ parameters for the $X$-state and with respect to $x$ and $r$
for Werner state. For Werner state, Fisher information decays as
the acceleration increases and the maximum contribution on the
total Fisher information comes from the classical part. This shows
that, the amount of classical information increases as the Unruh
acceleration increases and consequently the travelling state loses
its entangled  behavior and may turn into a separable state.
However, for the  $X$- state, the contributor part depends on the
estimated parameter. For example, if we estimate the $x$
parameter, we can see that the contribution from the pure part is
the largest one, which indicates that the number of the generated
pure state is large.

{\it In conclusion:} the precision of estimating the parameters of
the accelerated state depends on the degree of entanglement of the
initial accelerated system and the value of the acceleration.
Generally speaking the estimation precision decreases as Unruh
acceleration increases for Werner state, while for $X$-state,
there are  values of Unruh acceleration that maximize the
estimated value.

\end{document}